\documentclass[twocolumn,aps, prd]{revtex4}%
\usepackage{graphicx}
\usepackage{dcolumn}
\usepackage{bm}
\usepackage{amsmath}%
\usepackage{amsfonts}%
\usepackage{amssymb}
\begin{document}

\title{Gamma ray burst triggers at daytime and night-time interface}

\author{C. R. A. Augusto, J. B. Dolival, C. E. Navia, and K. H. Tsui}
\address{Instituto de F\'{\i}%
sica Universidade Federal Fluminense, 24210-130,
Niter\'{o}i, RJ, Brazil} 

\date{\today}
\begin{abstract}

There is a difference between the solar ionization concentration in the ionosphere in the daytime and at night-time. At night the E-region ion concentration peak is dramatically reduced due to chemical losses and the rapid change in the vertical polarization electric field at the time around sunset, which is due to the accelerating neutral wind dynamo and which produces a corresponding change in the zonal electric field through curl-free requirements. The result is the formation of a layer of high conductivity, at the daytime-night-time interface. This phenomenon in the South Atlantic Anomaly (SAA) area, provokes an increase in the precipitation of charge particles which is well known and is commonly termed ``sunset enhancement''. In the following we show five gamma ray burst (GRB) triggers observed by spacecraft GRB detectors in temporal coincidence with muon enhancement observed at ground level by the Tupi telescopes with two different orientations at $\sim 21$ UT (local sunset), and located inside the SAA region. Of these GRB triggers analyzed here, one from Swift, and two from Fermi are probably noise triggers, produced by omni-directional particle precipitation, during the sunset enhancements.

\end{abstract}


\maketitle

\section{Introduction}

Gamma ray bursts (GRBs) are a relatively newly identified transient phenomenon, discovered accidentally in the 1960's by American military satellites (Vela project) \cite{klebesabel73} Their discovery has brought us into a new, challenging phase of investigation. Most of the observed GRBs appear to be collimated emissions of gamma rays, observed in the deep sky at random places and at random times. Some of these bursts are followed by afterglows at X-ray, optical and radio wavelengths. Our present knowledge of GRB emissions comes from space-borne detectors in the keV to sub MeV energy band, some no longer in operation, such as  BATSE on board the CGRO \cite{hurley94,schonfelder96}, the Beppo-SAX and HETE. Presently in operation are the KONUS, INTEGRAL, AGILE, the multi band detector Swift and the Fermi. The description of the GRB phenomena above MeV is based mainly on the remarkable results obtained by EGRET on board CGRO (Hartman et al. 1999), and to make progress in this energy band two new gamma-ray experiments are in operation, namely the AGILE and Fermi.

The SAA is a region close to the west coast of Brazil in which the shielding effect of the magnetosphere is not perfectly spherical and it shows as a ``hole'' in the shield as a result of the eccentric displacement of the center of the magnetic field as compared with the origin (center of mass). This behavior of the magnetosphere is responsible for several processes, such as the trapped and azimuthally drifting energetic particles, bouncing between hemispheres, coming deeper down into the atmosphere owing to the low field intensity over SAA and thereby interacting with the dense atmosphere resulting in ionization and increasing electric conductivity.

In order to explore the potential of these favorable conditions, we have mounted two muon telescopes inside the SAA region. 
Also to be taken into account is the fact that the Earth's magnetic field deflects the charged particles of the shower initiated by primary cosmic rays including gamma rays. This deflection is caused by the component of the Earth's magnetic field perpendicular to the particle trajectory. This effect results in a decrease in the number of collected particles and therefore the telescope's sensitivity. This means that the sensitivity of particle telescopes is highest in the SAA region, because in this region the transverse magnetic field is very small, even smaller than the average magnetic field of the polar areas. 

So far, the results have been the observation of small transient solar events such as solar flares in association with the GOES satellite \cite{navia05,augusto05}. The precipitation of the charged particles in the SAA region, their fluctuations and some regularities such as the schedule of the precipitation, that in most of the cases begins at 09.00 UT and concludes at around the $\sim 21.00$ UT (that is our local sunset) along with a previous precipitation enhancement, the so called sunset enhancement. We have also shown that when space-borne GRB detectors are in or in close proximity to the SAA, the rate of charged particles is too high to be efficiently filtered out,  and trigger the GRB alarm of the detectors \cite{augusto08,augusto08b}. In some cases these noise triggers can be identified for the Tupi telescopes, because there are muon excess (peaks) in association with the GRB trigger occurrences, and the trigger coordinates are inside or close to the field of view of the telescopes. Finally, the main result obtained is a possible GeV counterpart observed at ground  of keV GRBs detected in space-borne detectors such as Swift and Fermi \cite{augusto08c,augusto08d}. 

In this paper we show that three recent GRB triggers, two from Fermi (GBM trigger), and one from Swift, they could, in fact, be noise triggers produced by particle precipitation during the sunset enhancement. This is because, in these cases, the trigger occurrences are at ($\approx 21$ UT) and coincide with our local sunset and, moreover, they are in association with muon enhancement: narrow peaks were observed with a high confidence level on the muon background in both telescopes. This means that the peaks are produced by an omni-directional radiation typical of the particle precipitation in the SAA region. In addition the trigger coordinates are close to the field of view of the Tupi telescopes. We also included in the analysis two recent GRB triggers, on from AGILE, and one from INTEGRAL space-borne GRB detectors.

The paper is organized as follows: in Section II, the Tupi experiment is described in brief. In order to understand the sunset enhancement, in Section III the daytime and night-time interface mechanism is briefly described. In Section IV the analysis of the GRB triggers is presented, and Section VI contains our conclusions.
 
\section{Experimental setup}

The Tupi experiment consists of the locating of two telescopes in the SAA region (22S and 43W). Starting in April of 2007 we initiated Phase II of the Tupi experiment with a survey of the daily variations in the muon intensity at sea level using the two identical muon telescopes. The method applied here to study the cosmic ray anisotropy is based on the idea that a fixed detector will scan the sky due to the Earth's rotation. Each Tupi telescope has two scintillator detectors and they are in coincidence. One of them has a vertical orientation, and the other is oriented at nearly 45 degrees to the vertical (zenith) pointing to the west as is shown in Fig.2. 
\begin{figure}[th]
\vspace*{-1.2cm}
\hspace*{-1.0cm}
\includegraphics[clip,width=0.55
\textwidth,height=0.6\textheight,angle=0.] {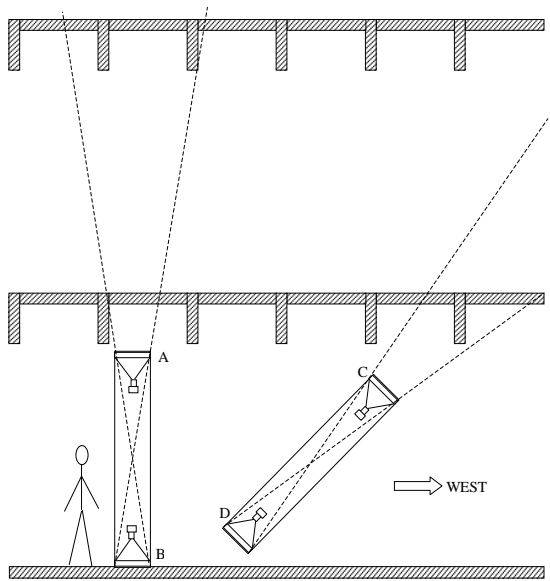}
\vspace*{-5.0cm}
\caption{The two muon telescopes of the Tupi experiment phase II}
\end{figure}
The telescopes are capable of detecting muons $(E_{\mu}>0.1 GeV)$ induced in the atmosphere by cosmic rays including gamma rays with primary energies above the pion production threshold ($\sim 10$ GeV). The directionality of the vertical muon telescope is guaranteed by a veto or anti-coincidence guard, using a detector of the inclined telescope and vice-versa. Thus, the incidence of an air shower activates the four detectors and is not registered and therefore only muons with trajectories close to the telescope axis are registered. Some details of the experimental setup were reported in \cite{augusto08}. Each telescope has an effective aperture of $0.27\;sr$.

\section{The daytime and night-time interface}

Fig.2 shows schematically the ion conductivity distribution of the F-region and E-region of the Earth's ionosphere as a function  of the altitude for typical daytime and nighttime conditions forming a tube \cite{heelis04}.
The ion number densities in the Earth's ionosphere, during the daytime have two peaks, one in the E-region 
and another in the F-region. These peaks are produced by solar ionization balanced
by chemical losses and diffusion. At night the E-region peak is dramatically reduced by the chemical losses while
downward diffusion serves to maintain the F-region peak. 
\begin{figure}[th]
\vspace*{-6.2cm}
\hspace*{-2.7cm}
\includegraphics[clip,width=0.8
\textwidth,height=0.8\textheight,angle=0.] {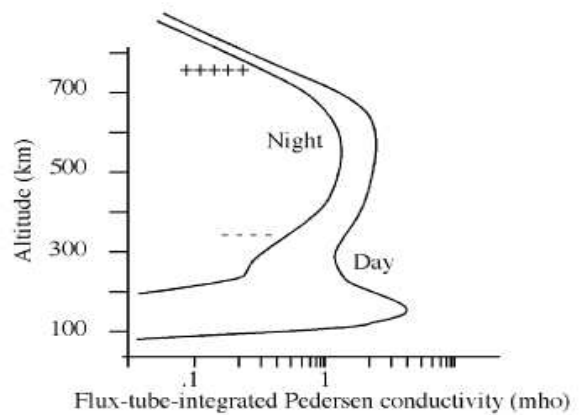}
\vspace*{-7.0cm}
\caption{The ionosphere conductivity variation with atmospheric altitude, for daytime and night-time conditions forming a tube (from ref.\cite{heelis04})}.
\end{figure}
A pre-reversal enhancement of the zonal electric field is produced due to rapid change in the polarization of the vertical electric field near to sunset, which is due to the accelerating neutral wind dynamo. Signs on the night-time profiles indicate where polarization charges will accumulate to create the electric field and to produce a $E \times B$ zonal drift of the charged particles in the F-region
\begin{figure}[th]
\vspace*{-4.2cm}
\hspace*{-3.0cm}
\includegraphics[clip,width=0.8
\textwidth,height=0.8\textheight,angle=0.] {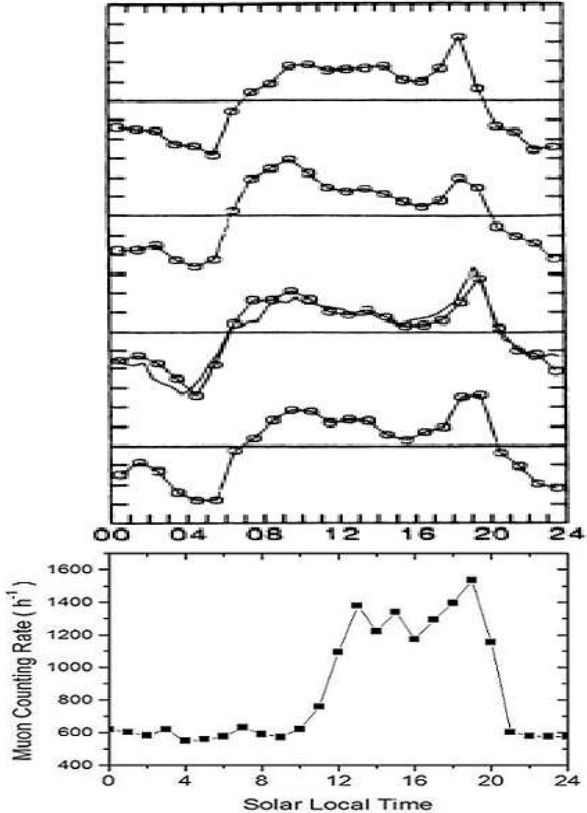}
\vspace*{-4.0cm}
\caption{(Upper panel) Empirical model of satellite-measured vertical drifts under magnetically quiet conditions (Nov-Feb) \cite{fejer95}, and for four longitudinal sectors with averages $80^0$, $180^0$, $260^0$ and $320^0$. (Lower panel) Average hourly muon counting rate observed in March 2008 by the Tupi vertical telescope.}
\end{figure}

The vertical electric field and vertical drifts were investigated particularly, in order to relate the electro-dynamic processes at sunset to an existing empirical model of satellite experimental results on vertical drift. We have compared the average hourly muon counting rate observed in March 2008 (minimal solar activity) and the empirical model of satellite measured vertical drift under magnetically quiet conditions \cite {fejer95}. The results are shown in Fig.3. We would like to make two comments: firstly that in all cases the sunset enhancement ($\sim 18h$ local time) can be observed with a high level of confidence. Secondly, that the vertical drift enhancement begins at sunrise ($\sim6h$ local time), while the charge particle precipitation is  observed only 3 hours after sunrise and continues after sunset, because the interplanetary magnetic field is overtaking the Earth at these hours. The SAA region can be considered as an ``open magnetosphere'' and the so-called "reconnection magnetic process" takes place in the SAA region 3 hours after sunrise and produces field lines with one end at the Earth and the other in distant space.

\section{The GRB triggers}

So far, the following five GRB trigger occurrences have been observed practically in association with ground level enhancements (GLEs). That is to say, muon excess is observed as narrow peaks (with a high confidence level) against the muon background level observed by the Tupi telescopes at sea level. 
We argue here that these triggers (except for the AGILE and INTEGRAL triggers) have a great chance of being classifiable as noise triggers 
produced by an omni-directional radiation during the sunset enhancement and this is a characteristic of the precipitation of particles in the SAA region. The triggers are presented in chronological order. 

\subsection{Fermi-GBM trigger 241307335/080824909}

Fermi is an interesting project dedicated to gamma-ray astrophysics \cite{gargano08}. The Fermi has a Gamma Burst Monitor (GBM) which works to support the Large Area Telescope (LAT) in observing GRBs through providing low-energy measurements with a high time resolution and also by providing rapid burst locations over a large field of view ($\geq 8$ sr). The GBM will complement the LAT (energy range: 10 MeV to $> 100$ GeV) measurements by observing GRBs in the energy range of 10 keV to 30 MeV. About fifteen times per month there will be a GRB in the field of view. Independently of this, the LAT will also detect about two bursts per month.

\begin{figure}[th]
\vspace*{-0.0cm}
\includegraphics[clip,width=0.4
\textwidth,height=0.4\textheight,angle=0.] {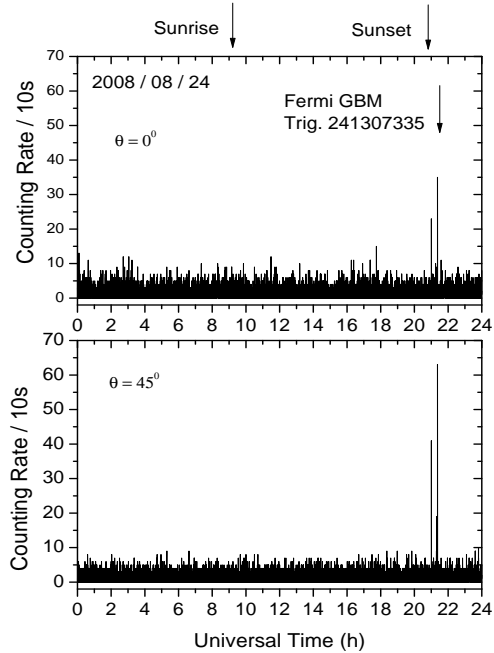}
\vspace*{-0.0cm}
\caption{The muon counting rate at $1/10\; s^{-1}$ observed on August 24, 2008 in the vertical (upper panel) and in the inclined (lower panel) muon telescopes. The vertical arrow indicates the occurrence of the FERMI-GBM trigger 241307335.}
\end{figure}
According to GCN CIRCULAR 8144 \cite{wilson}, on 2008 August 24 at 21:48:54 UT the Fermi-GBM alert reported the arrival of a GRB, Trigger 241307335/080824909. The on-ground calculated location, using the Fermi GBM trigger data, was (RA, DEC) = (120.3d, -3.2d), the event is classified as a GRB with a single peak and tail, with a duration of $T90=28s$ and fluence of $2.3\times 10^ {-6} erg/cm^2$ in the 500-300 keV energy band. The trigger occurrence took place 17 minutes after a very narrow muon peak was observed in both telescopes. The situation is summarized in Fig.4. As the peaks are in both the two telescopes, we can conclude that an omni-directional precipitation of particles is at the origin of the peaks. They have a confidence level of $20\sigma$ in the vertical telescope and $31\sigma$ in the inclined telescope. A plentiful precipitation was not observed on this day, the precipitation with a high intensity only happened just at sunset.
\begin{figure}[th]
\vspace*{-1.0cm}
\includegraphics[clip,width=0.5
\textwidth,height=0.5\textheight,angle=0.] {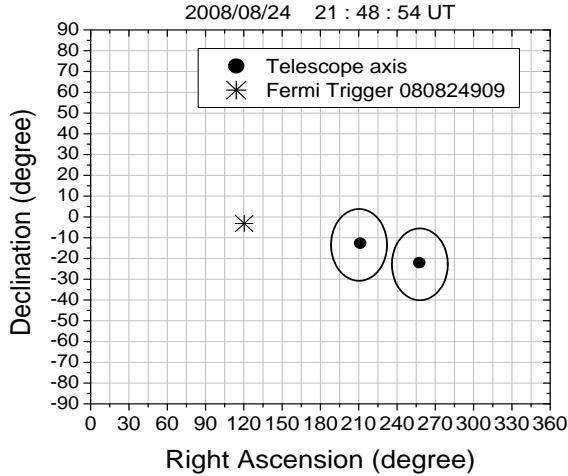}
\vspace*{-4.7cm}
\caption{Equatorial coordinates, showing the position of the two Tupi telescope axes at GRB trigger time, as well as the Fermi GBM 080824909 coordinates. The ``ellipses'' represent the effective field of view (0.27 sr) of the Tupi vertical telescope.}
\end{figure}

However, the GRB trigger coordinates are a little far from the field of view of the two telescopes as is shown in Fig.5. This probably means that drift processes in the particle precipitations at sunset are responsible for this trigger. In addition, this Fermi trigger is not part of the official Fermi GBM Trigger catalog \cite{fermib}.

\subsection{AGILE GRB trigger 691}

AGILE (Astro-rivelatore Gamma a Immagini LEggero) is a Scientific Mission
dedicated to high-energy gamma-ray astrophysics supported by the Italian Space Agency (ASI) \cite {pittori03}. Super-AGILE will be able to locate GRBs within a few arcminutes; AGILE has excellent gamma-ray imaging with a large Field-of-View (FOV) which is $\sim 3\; sr$, larger than EGRET by a factor of $\sim 4-5$, and it has an alert system.
\begin{figure}[th]
\vspace*{-1.0cm}
\includegraphics[clip,width=0.5
\textwidth,height=0.5\textheight,angle=0.] {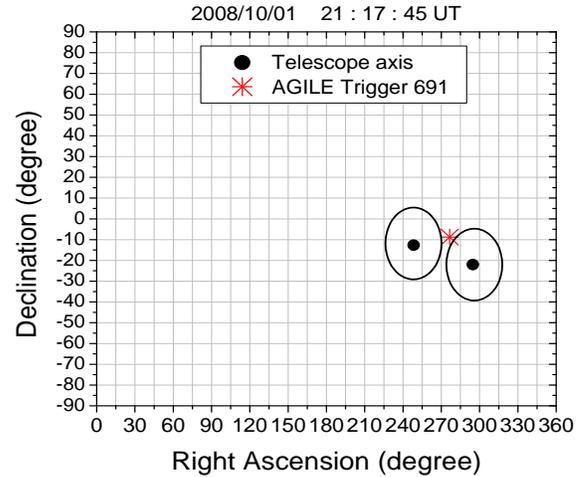}
\vspace*{-5.0cm}
\caption{Equatorial coordinates, showing the position of the two Tupi telescope axes at GRB trigger time, as well as the AGILE GRB trigger 691 coordinates. The ``ellipses'' represent the effective field of view (0.27 sr) of the Tupi vertical telescope.}
\end{figure}

The AGILE Gamma-Ray Imaging Detector (GRID) is sensitive in the 30MeV-50GeV energy range. Additional hard X-ray detection capability is provided by the Super-AGILE detector (SA). The CsI Mini-Calorimeter (MCAL) will also detect events independently of the GRID. The energy range for this non-imaging detector is $0.3-200MeV$ and will provide spectral and accurate timing information for transient events.

\begin{figure}[th]
\vspace*{-0.5cm}
\hspace*{-1.0cm}
\includegraphics[clip,width=0.6
\textwidth,height=0.4\textheight,angle=0.] {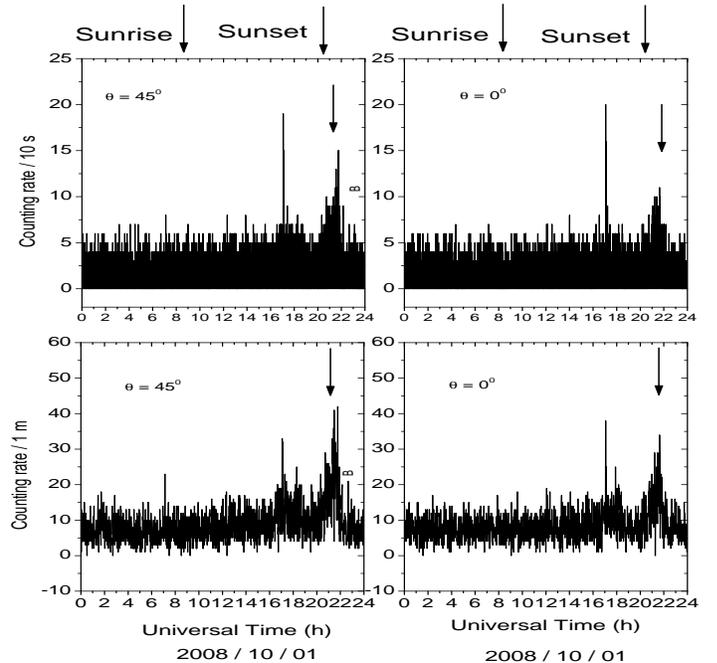}
\vspace*{-0.8cm}
\caption{The muon counting rate at $1/10\; s^ {-1} $ (upper) and at $1/60\; s^ {-1} $ (lower) observed on October 1, 2008 in the inclined (left panels) and vertical (right panels) muon telescopes. The vertical arrow indicates the occurrence of the AGILE GRB trigger.}
\end{figure}

One of the most interesting associations between muon excess and GRB triggers, is one that was observed with AGILE GRB trigger 691 in effect the GRB trigger coordinates also lie inside the field of view of the two telescopes as is shown in Fig.6. The trigger occurrence was at 21:17:45 on October 1, 2008 and coincides with a copious omni-directional particle precipitation during the sunset. The precipitation was observed at the counting rate of every 10 s by both Tupi telescopes with a confidence of $8.1\sigma$ in the inclined and $5.7\sigma$ in the vertical telescopes respectively, as shown in Fig.7 upper panels. The confidence level is further enhanced by $\sim 100\%$ in the counting rate at every minute, as is shown in the lower panels of Fig.7. According to the GCN CIRCULAR 8305 \cite{lazzarotto}, the AGILE event had a duration of about 40 s in the 20-60 keV energy range, with a first 10 s long bright peak, followed by a second broad peak separated by a break of about 20 s.

The situation requires care, since in this case the AGILE GRB coordinates are practically simultaneously inside the field of vision of the two telescopes and the possibility exists of this being a true GRB, according to the GCN CIRCULAR 8307 \cite{page} this AGILE trigger has been confirmed as a GRB by the observation of its X-ray afterglow with Swift, four hours after the burst. In addition the behavior of imaging in the detector is consistent with a beam of photons.
  
\subsection{INTEGRAL-IBAS GRB trigger 5361}

The International Gamma Ray Astrophysics Laboratory (INTEGRAL) is a gamma-ray astronomy satellite belonging to the European Space Agency \cite {winkler96}. The INTEGRAL spacecraft can detect and locate GRB thanks to the large field of view, unprecedented angular resolution and the high sensitivity of the IBIS imaging instrument (20 keV-10 MeV). The GRB coordinates with an accuracy of a few arcmin, and it is distributed within 2-3 minutes from the time of GRB detection on the satellite, by means of the INTEGRAL Burst Alert System (IBAS). The gamma ray burst coordinates network (GCN) system also incorporates the distribution of locations of GRBs detected by the INTEGRAL spacecraft; both the localizations by ISGRI and the unlocalized detections by the SPI Anti-Coincidence Shield (ACS).
\begin{figure}[th]
\vspace*{-0.5cm}
\hspace*{-1.0cm}
\includegraphics[clip,width=0.6
\textwidth,height=0.4\textheight,angle=0.] {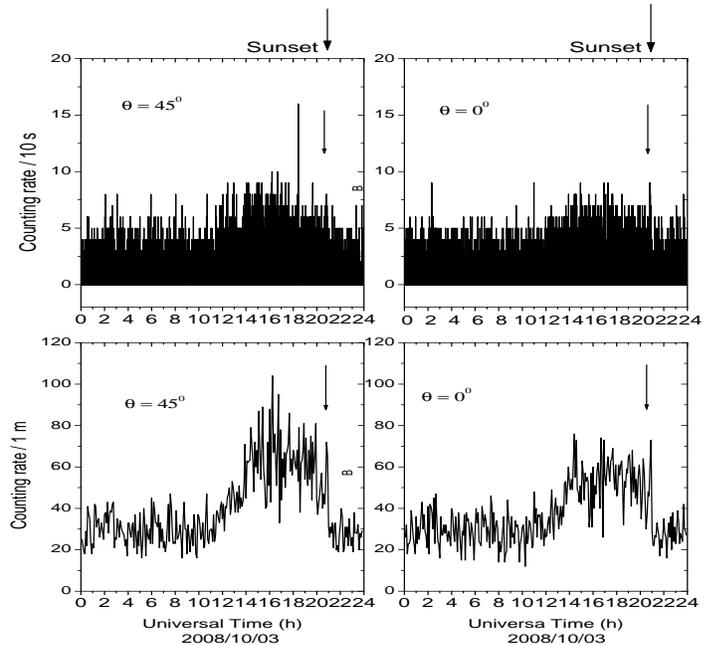}
\vspace*{-1.0cm}
\caption{The muon counting rate at $1/10\; s^ {-1} $ (upper) and at $1/60\; s^ {-1} $ (lower) observed on October 3, 2008 in the inclined (left panels) and vertical (right panels) muon telescopes. The vertical arrow indicates the occurrence of the INTEGRAL-IBAS GRB trigger.}
\end{figure}
According to the GCN CIRCULAR 8317 \cite{gotz08}, on October 3, 2008 at 20:48:17 UT the Integral Burst Alert System (IBAS) on board of the INTEGRAL spacecraft reported the arrival of a GRB, Trigger 5361. The event is classified by INTEGRAL as a possible GRB event. A GRB lasting about $30$ seconds, and a peak in the energy band of 20-200 keV. The IBAS event Position was (RA: DEC) = (273.5d: +17.50d). 
\begin{figure}[th]
\vspace*{-1.2cm}
\includegraphics[clip,width=0.5
\textwidth,height=0.5\textheight,angle=0.] {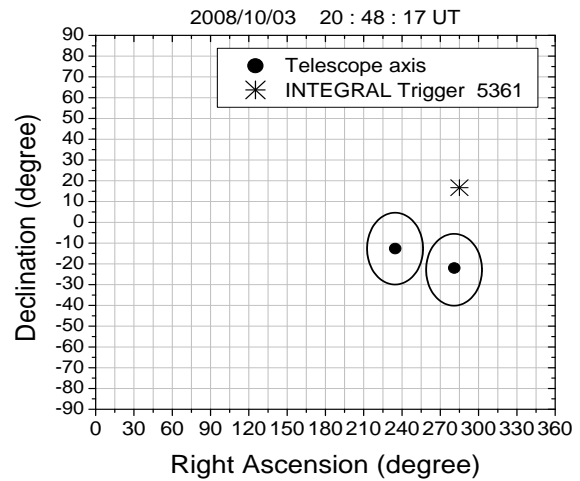}
\vspace*{-4.6cm}
\caption{Equatorial coordinates, showing the position of the two Tupi telescope axes at GRB trigger time, as well as the INTEGRAL-IBAS trigger 5361 coordinates. The ``ellipses'' represent the effective field of view (0.27 sr) of the Tupi vertical telescope.}
\end{figure}

Figure 8 (upper panels) shows the telescope output (raw data) representing the 10 second muon counting rate for the inclined and vertical telescopes respectively. It is possible to see that the INTEGRAL trigger occurrence (vertical arrow) was during a copious particle precipitation in the SAA region and that it coincides with the sunset enhancement observed with a confidence level of $5.0\sigma$ in the inclined telescope. This signal is more intense in the vertical telescope ($5.4\sigma$). In addition, the confidence level is enhanced by $\sim 60\%$ in the counting rate every minute, as is shown in the lower panels of Fig.8. Moreover, the coordinates of this INTEGRAL trigger are close to the effective field of view of the Tupi telescopes as is shown in Fig.9, where the two telescopes' axes of equatorial coordinates are indicated together with the GRB coordinates. 

However, this INTEGRAL trigger is a real possible GRB, because the INTEGRAL spacecraft has a highly elliptic orbit wit a period of 3 days, 150000 km of apogee and 10000 km of perigee and during the trigger occurrence, the INTEGRAL spacecraft was at an height of about 100000 km, above but very far from the SAA region. In addition the behavior of imaging in the detector is consistent with a beam of photons.

\section{Swift trigger 338633 and Fermi trigger bn081230871}

Swift is a multi-wavelength space-based observatory dedicated to the study of gamma-ray burst (GRB) science. Its three instruments work together to observe GRBs and their afterglows in the gamma-ray, X-ray, ultraviolet, and optical wavebands. The satellite was developed by an international consortium from the United States, United Kingdom, and Italy \cite{gehrels04}.
The Burst Alert Telescope (BAT) is a 15 - 150 keV energy range detector, and it detects a new GRB event and computes its coordinates in the sky.
\begin{figure}[th]
\vspace*{-0.5cm}
\includegraphics[clip,width=0.4
\textwidth,height=0.4\textheight,angle=0.] {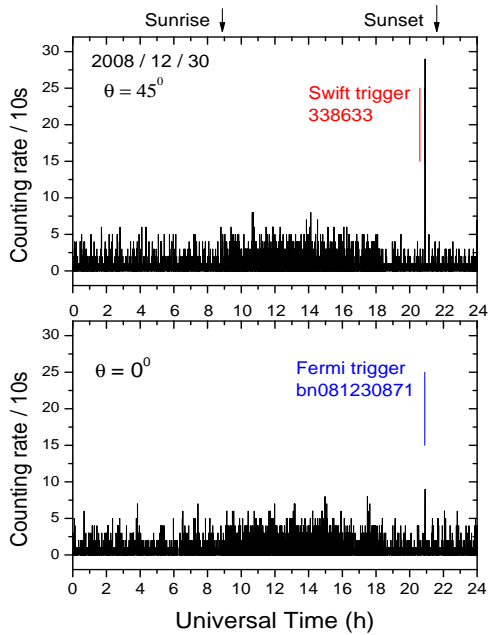}
\vspace*{-0.4cm}
\caption{The muon counting rate at $1/10\; s^{-1}$ observed on December 30, 2008 in the inclined (upper panel) and in the vertical (lower panel) muon telescopes. The vertical red and blue lines indicates the occurrence of the Swift trigger 338633 and FERMI-GBM trigger bn081230871 respectively.}
\end{figure}

Figure 10 shows the telescope output (raw data) on December 30, 2008 and it representing the 10 second muon counting rate for the inclined (upper panel) and vertical (vertical panel) telescopes respectively. A plentiful precipitation was not observed on this day, the precipitation with a high intensity only happened just at sunset. The red and blue vertical lines indicates the occurrence of the Swift trigger 338633 and Fermi trigger bn081230871 respectively. They are practically in temporal coincidence com a very sharp peak in the muon counting rate and close to the sunset. In addition the coordinates of the Swift trigger are very close to the effective field of view of the inclined telescope, and the coordinates  of the Fermi trigger are inside of the effective field of view of the inclined Tupi telescope. The figures 11 and 12 summarize the situation. 
\begin{figure}[th]
\vspace*{-1.6cm}
\includegraphics[clip,width=0.5
\textwidth,height=0.5\textheight,angle=0.] {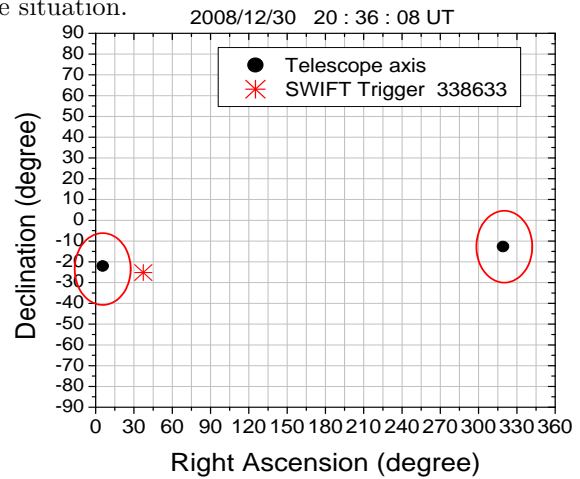}
\vspace*{-5.0cm}
\caption{Equatorial coordinates, showing the position of the two Tupi telescope axes at GRB trigger time, as well as the Swift trigger 338633 coordinates. The ``ellipses'' represent the effective field of view (0.27 sr) of the Tupi vertical telescope.}
\end{figure}

\begin{figure}[th]
\vspace*{-1.3cm}
\includegraphics[clip,width=0.5
\textwidth,height=0.5\textheight,angle=0.] {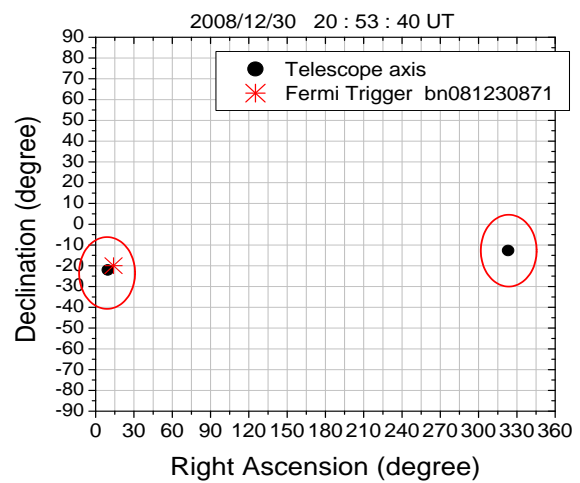}
\vspace*{-5.0cm}
\caption{Equatorial coordinates, showing the position of the two Tupi telescope axes at GRB trigger time, as well as the Fermi GBM trigger bn081230871 coordinates. The ``ellipses'' represent the effective field of view (0.27 sr) of the Tupi vertical telescope.}
\end{figure}

Initially the Swift trigger was classified as a GRB \cite{GCN_SWIFT_NOTICE}. However, in the posterior GCN NOTICE DATA has been indicated that the XRT position is 0.93 arc-min from BAT position. The object found at this position is either a very bright burst or a cosmic ray hit and finally the Swift SC slew decision it has not been confirming this trigger as being GRB. The situation of the Fermi trigger bn081230871 is more delicate, because in the Fermi GBM Burst Catalog
(fermigbrst) \cite{fermib} the probability of the trigger classification as being a GRB is 0.9882.
In spite of this conclusion, there are three arguments that indicate that this trigger has chance of being a noise trigger. (a) the trigger is practically in temporal coincidence with the muon enhancement at sunset, (b) the coordinates of the trigger are inside of the field of view of the inclined telescope, and (c) exactly  it is in the inclined telescope that the confidence level of the muon enhancement is bigger ($24.6\sigma$), the confidence level in the vertical telescope is only   
$5,2\sigma$.

\section{Conclusions} 

Since April 2007, we have been operating phase II of the Tupi experiment, with two directional muon telescopes located at sea level at 22S and 43W. These coordinates are within the SAA region and close to its center. This characteristic provides the muon telescopes with very low rigidity of response to cosmic protons and ions ($\geq 0.4$ GV). We have been monitoring the particle precipitations in the SAA region with a high duty cycle (above 95\%). We believe that monitoring the precipitation of particles continuously can be useful for the identification of some GRB triggers, as well as for the understanding of the mechanisms of the precipitation of particles; their regularities and fluctuations, because they are subject to seasonal variations.

We would like to point out that the Swift triggers observed on January 24, 2008  have been confirmed  as noise triggers in the  CGN CIRCULAR NUMBER 7213 \cite{barthelmy}. These triggers occurred while Swift was entering the South Atlantic Anomaly (SAA). We have shown that these Swift triggers are associated with Tupi muon peaks produced by the high energy particle precipitation in the SAA region \cite{augusto08}
The three GRB triggers one from Swift, and two from Fermi analyzed here have similar characteristics:

(a) The trigger occurrences are very close to sharp muon peaks and occur inside or not far from the field of view of the telescopes. 

(b) The muon peaks are observed in both telescopes. This means that the enhancement is produced by an omni-directional radiation which is a characteristic of the precipitation of particles in the SAA region.

We would like pointed out, that the AGILE and INTEGRAL triggers analyzed here have similar characteristic. However, there are several reasons, such as the observation of X-ray afterglow in the case of AGILE GRB and the highly elliptical orbit in the case of INTEGRAL spacecraft
demonstrate that AGILE and INTEGRAL triggers cannot be due to particle precipitation in the SAA. Consequently, these two triggers
occurs independently, the coincidence with the sunset enhancement in the SAA region is only fortuitous.

In all cases, they occur at daytime and night-time interface, a time when a highly conductive vertical layer forms in the atmosphere which initiates the precipitation of the charged particles; the so-called "sunset enhancement". Consequently, the rate of particles at sunset is too high to be efficiently filtered out by the space-borne GRB detectors.

The Tupi experiment is still underway and in a process of expansion. We will put fourteen more telescopes into operation within six months, increasing the field of view of the Tupi experiment sixteen-fold.

\acknowledgements

This work is supported by the Brazilian National Council for Research (CNPq), under Grant No. $479813/2004-3$ and
$476498/2007-4$. We are also grateful to the various catalogs available  on the web and to
their open data policy, especially to the GCN report. We want to give special gratefulness to Sandro Mereghetti for the valuable informations on the AGILE and INTEGRAL triggers.


\newpage

\begin{references}
\bibitem{klebesabel73}R. W. Klebesabel, I. B. Strong, and R. A. Olson,  Astrophys. J. 182, L85 (1973).

\bibitem{hurley94}K. Hurley et al., Nature 372, 652 (1994).
\bibitem{schonfelder96}V. Schonfelder, Il Nuovo Cimento, 19, 805 (1996).


\bibitem{navia05}C. E. Navia, C. R. A. Augusto, M. B. Robba, M. Malheiro and H. Shigueoka, Astrophys. J. 621, 1137 (2005).
\bibitem{augusto05} C. R. A. Augusto, C. E. Navia and M. B. Robba, Phys. Rev. D71, 103011 (2005).

\bibitem{augusto08}C. R. A. Augusto, C. E. Navia and K. H. Tsui, Phys. Rev. D 77, 123008 (2008).
\bibitem{augusto08b}C. R. A. Augusto, C. E. Navia and K. H. Tsui, Phys. Rev. D 78, 087102 (2008).

\bibitem{augusto08c}C. R. A. Augusto, C. E. Navia, M. B. Robba, and K. H. Tsui, Phys. Rev. D 78, 122001 (2008).
\bibitem{augusto08d}C. R. A. Augusto, C. E. Navia and K. H. Tsui, and H. Shigueoka. astro-ph/0812.1935

\bibitem{heelis04}R. A. Heelis, Journal of Atmospheric and Solar-Terrestrial Physics, 66, 825 (2004)

\bibitem{fejer95}B. G. Fejer and L. Scherlies, Geophys. Research Lett. 22, 851 (1995).

\bibitem{gargano08}F. Gargano, 17th International Workshop on Vertex detectors July 28 – August 1, 2008, Utö Island, Sweden, available at $http://glast2.pi.infn.it/SpBureau/glast-lat-contribution-2/vertex-2008-contribution/talk.2008-07-10.1022714254/at-down
load/file$

\bibitem{wilson}C. A. Wilson, GCN CIRCULAR 8144, available at:\\
$http://gcn.gsfc.nasa.gov/gcn3/8144.gcn3$

\bibitem{fermib}Fermi GBM Trigger Catalog, available at:\\
$http://heasarc.gsfc.nasa.gov/cgi-bin//W3Browse/w3query.pl$


\bibitem{pittori03}C. Pittori et al., Chin. J. Astron. Astrophys. Vol. 3 (2003), Suppl., 517–522.

\bibitem{lazzarotto} F. Lazzarotto et al., GCN CIRCULER 8305, available at:\\
$http://gcn.gsfc.nasa.gov/gcn3/8305.gcn3$

\bibitem{page}K. Page et al., GCN CIRCULAR 8307, avilable at:\\
$http://gcn.gsfc.nasa.gov/gcn3/8307.gcn3$

\bibitem{winkler96}C. Winkler, Astron. Astrophys. Suppl. Ser. 120, pp. 637-640 (1996).
\bibitem{gotz08} D. Gotz et al.,GCN CIRCULAR 8317 available at:\\
$http://gcn.gsfc.nasa.gov/gcn3/8317.gcn3$

\bibitem{gehrels04} N. Gehrels et al., \apj, 611, 1005 (2004).

\bibitem{GCN_SWIFT_NOTICE}GCN SWIFT NOTICE, available at:\\
$http://gcn.gsfc.nasa.gov/other/337115.swift$

\bibitem{barthelmy}S. Barthelmy, GCN CIRCULAR 7213, available at:\\
$http://gcn.gsfc.nasa.gov/gcn3/7213.gcn3$



\end{references}
\end{document}